\documentclass[prl,aps,showpacs,twocolumn,unsortedaddress]{revtex4}
\usepackage{graphics,bm}
\usepackage{amssymb}
\usepackage{epsfig}
\usepackage{epsf}

\begin{document}

\title{Itinerant Ferromagnetism in an Ultracold Atom Fermi Gas}

\author{R.A. Duine}
\email{duine@physics.utexas.edu}
\homepage{http://www.ph.utexas.edu/~duine}

\author{A.H. MacDonald}
\email{macd@physics.utexas.edu}
\homepage{http://www.ph.utexas.edu/~macdgrp}

\affiliation{Department of Physics, University of Texas at Austin,
1 University Station C1600, Austin, TX 78712-0264}

\date{\today}

\begin{abstract}
We address the possible occurrence of ultracold
atom ferromagnetism by evaluating the free energy of a spin polarized
Fermi gas to second order in its interaction parameter.
We find that Hartree-Fock theory underestimates the
tendency toward ferromagnetism, predict that the ferromagnetic transition
is first order at low temperatures, and point out that the spin
coherence time of gases prepared in a ferromagnetic state is
strongly enhanced as the transition is approached. We relate our
results to recent experiments.
\end{abstract}

\pacs{03.75.Ss, 71.10.Ca, 32.80.Pj}

\maketitle

\def\bx{{\bf x}}
\def\bk{{\bf k}}
\def\half{\frac{1}{2}}
\def\args{(\bx,t)}

\noindent {\it Introduction} --- Itinerant ferromagnetism is
common in metals.  Nevertheless, because it flows from a
strong-coupling Fermi liquid instability, the microscopic physics
that controls its occurrence is less well understood than the
physics that controls superconductivity \cite{shankar}. In the
electron gas case, for example, accurate quantum Monte Carlo
calculations suggest \cite{montecarlo} that the transition to the
ferromagnetic state occurs at a critical density nearly three
orders of magnitude smaller than that predicted by mean-field
(Hartree-Fock) theory. Even in the simplest model of interacting
electrons, the single-band Hubbard model, solid predictions on the
occurrence of ferromagnetism are rare and often restricted to
particular band fillings \cite{vollhardt,bookauerbach}.
Understanding the nature of the paramagnetic to ferromagnetic
phase transition, when it occurs, has also been challenging.
Experimental progress has recently been achieved by applying
hydrostatic pressure to itinerant ferromagnets with a low Curie
temperature, making it possible to study the transition in the
zero temperature limit and test for theoretically predicted
quantum critical \cite{hertz1976} behavior. In these experiments,
the line of continuous transitions in the temperature-pressure
phase diagram appears \cite{pfleiderer1997,saxena2000,nicklas1999,
uhlarz2004} to terminate at a tricritical point with decreasing
temperature, connecting with a low-temperature line of first order
transitions. In mean-field theory, first order magnetic
transitions can follow from a non-monotonic quasiparticle
density-of-states near the paramagnetic state's Fermi energy
\cite{shimizu1964}. Belitz {\it et al.}
\cite{belitz1999} have argued, however, that coupling of the order parameter
to gapless modes leads to nonanalytic terms in the free energy,
and generically drives the transition first order. These
nonanalytic terms were first predicted by Misawa on the basis of
Fermi-liquid theory \cite{misawa1971}, and are a consequence of
gapless particle-hole excitations. Theories of the phase
transition are still qualitative however, and detailed
experimental corroboration of this picture is still lacking.

In this Letter we address the possible complementary realization
of ferromagnetism in ultracold fermionic atoms, which are
accurately described by a short-range interaction model
\cite{bookpathria,kanno1970}. In Hartree-Fock theory
\cite{bookpathria} the zero-temperature ferromagnetic transition
of this model is continuous and the ground state is ferromagnetic
when the gas parameter, {\em i.e.}, the product of the Fermi wave
number $k_{\rm F}$ of the unpolarized system, and the $s$-wave
scattering length $a$ of the short-range potential, satisfies
$k_{\rm F} a \geq \pi/2$.  The phase separation predicted by
Houbiers {\it et al.} \cite{houbiers1997} at the same gas
parameter is one plausible manifestation of ferromagnetism but, as
we discuss below, not the most likely one. Trapped-atom motivated
inhomogeneous generalizations of these Hartree-Fock theories have
recently been analyzed by Salasnich {it et al.}
\cite{salasnich2000} and Sogo and Yabu \cite{sogo2002}.

The issue of ferromagnetism in a two-component atomic Fermi gas is
of particular interest because of the ongoing experimental study
of strongly interacting, degenerate, fermionic alkali atoms
\cite{demarco2001,truscott2001,schreck2001,granade2002,jochim2002,roati2002,hadzibabic2003}.
The focus so far has been on observing the formation of a fermion
pair condensate
\cite{strecker2003,cubizolles2003,jochim2003,greiner2003,zwierlein2003,regal2004,zwierlein2004,chin2004,gupta2003,regal2003}
in the BCS-BEC crossover
\cite{eagles1969,leggett1980,nozieres1985} regime close to a
Feshbach resonance \cite{stwalley1976,tiesinga1993}. Our interest
is in the repulsive interaction side of the resonance, where we
believe it will be possible to achieve unprecedented experimental
control over ferromagnetism. In making this assertion we are
assuming that the formation time of the molecular BEC state (which
occurs under the same conditions when the state is prepared by
crossing from the attractive interaction side of the resonance)
can exceed experimental time scales when the state is prepared by
approaching the resonance from the repulsive interaction side.

The character of the ferromagnetic state that can be realized
experimentally in these systems requires some comment
\cite{nygaard}. Since $s$-wave scattering does not occur between
identical fermions, interaction effects require the presence of
two hyperfine (pseudospin) species. Using standard techniques, the
atomic system can be prepared in a pseudospin coherent
(ferromagnetic) state, in which all atoms share the same spinor:
\begin{equation}
\label{eq:fullypolarizedstate}
  \left| \Psi_{\rm FM} (t) \right. \rangle =
  \frac{1}{\sqrt{2}} \prod_{|\bk|<2^{\frac{1}{3}} k_{\rm F}}
  \left( c_{\bk,\uparrow}^\dagger+e^{i (\varphi - \Delta E t/\hbar) }
   c_{\bk,\downarrow}^\dagger
  \right)|{\rm vac} \rangle~.
\end{equation}
($c_{\bk,\alpha}^\dagger$ creates an atom with momentum $\bk$ and
hyperfine spin $\alpha$.)  In Eq.~(\ref{eq:fullypolarizedstate}),
$\varphi$ specifies the orientation of the magnetic order
parameter in the $x-y$ plane and $\Delta E$ is the Zeeman energy
difference between the hyperfine states. (Since the number of
atoms in each is conserved, we can transform to a rotating wave
picture and let $\Delta E \to 0$.) Overall spin polarizations in
the $\hat{z}$ direction are not accessible. This fully spin
coherent state always has a lower energy than the phase-separated
state discussed in
Refs.~\cite{houbiers1997,salasnich2000,sogo2002} since, in the
magnetic language, the latter has a domain wall which costs finite
energy. Ferromagnetism in these systems will be manifested by
persistent coherence between hyperfine states.

\begin{figure}
\includegraphics{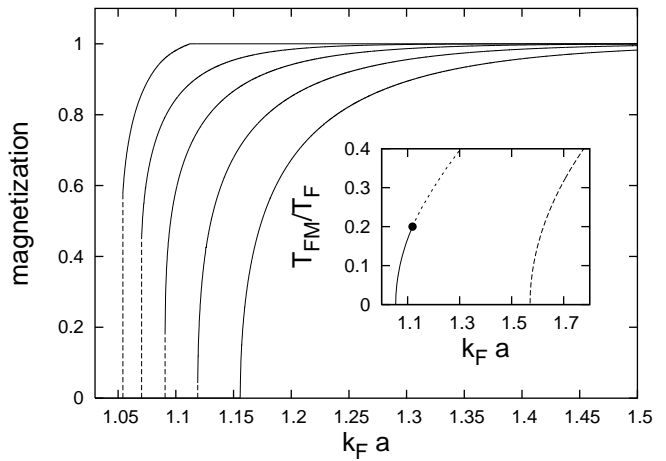}
\caption{\label{fig:combined} Magnetization $\xi$ as a function of
$k_{\rm F} a$, for various temperatures. From left to right
$T/T_{\rm F}=0,0.1,0.15,0.2,0.25$. The dashed lines indicate
magnetization jumps. The inset shows the critical temperature as a
function of the gas parameter. The solid line indicates
first-order transitions, and the dotted line second-order
transitions. The dashed line is the Hartree-Fock theory result.}
\end{figure}

\begin{figure}
\includegraphics{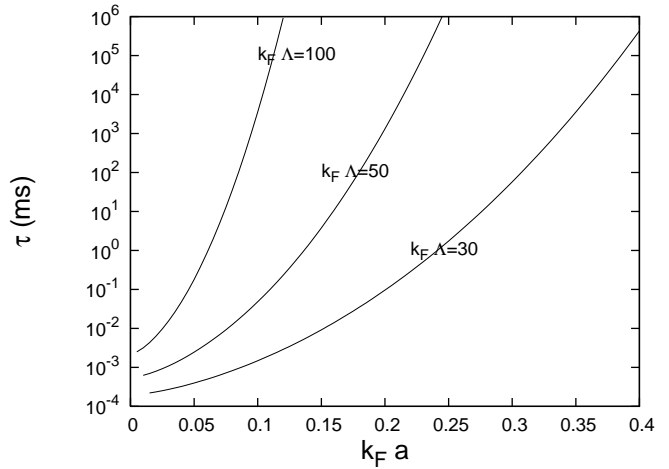}
\caption{\label{fig:suppression} Magnetic field inhomogeneity limit
on the spin coherence time of the
fully-polarized state.}
\end{figure}

In this Letter we argue that ferromagnetism occurs on the
repulsive interaction side of a Feshbach resonance. Our principle
results are summarized in Fig.~\ref{fig:combined} and
Fig.~\ref{fig:suppression}. We find that \emph{i}) Hartree-Fock
theory underestimates the tendency towards ferromagnetism
\cite{footnote1}, \emph{ii}) that the transition between
ferromagnetic and paramagnetic states is first order at low
temperatures and, \emph{iii}) that the coherence decay rate
decreases rapidly as the thermodynamic stability region of the
ferromagnetic state is approached from the repulsive side of the
resonance.

\noindent {\it Second order perturbation theory} --- It is
convenient to view the gas as a mixture of two independent
noninteracting gases of spinless fermions. The grand-canonical
Hamiltonian of the system is then
\begin{eqnarray}
\label{eq:hamiltonian}
  H = \int\!d\bx \sum_{\alpha=\left\{
+,-\right\}} \psi^\dagger_\alpha (\bx)\left(- \frac{\hbar^2 {\bf
\nabla}^2}{2m}-\mu_\alpha \right) \psi_\alpha (\bx)  \nonumber \\
+ g \int\!d\bx\psi^\dagger_+ (\bx) \psi^\dagger_- (\bx)
 \psi_- (\bx) \psi_+ (\bx)~,
\end{eqnarray}
with $g=4 \pi a \hbar^2/m$. The chemical potentials are determined
by $n_\alpha = \partial p_{0\alpha}/\partial \mu_\alpha$, where
$n_\alpha$ is the density of atoms in hyperfine state $|\alpha
\rangle$, and the pressure of the noninteracting gas is given by
\begin{equation}
  p_{0\alpha} = \frac{k_{\rm B} T}{V} \sum_\bk \ln \left[
1+e^{-\beta(\epsilon_\bk - \mu_\alpha)} \right]~,
\end{equation}
with $k_{\rm B} T$ the thermal energy, $V$ the volume, and
$\epsilon_\bk = \hbar^2 \bk^2/2m$ the single-particle dispersion.
The entropy density is determined by $s=\partial
(p_{0+}+p_{0-})/\partial T$, and the total free energy density is
given by $f(n_+,n_-)=e-Ts$, with the total energy density expressed as the sum
of three contributions, $e=e^{(0)}+e^{(1)}+e^{(2)}$.
The first two contributions correspond to Hartree-Fock theory,
and are given by
\begin{equation}
 e^{(0)}+e^{(1)} =
  \frac{1}{V} \sum_\bk \left[  \sum_{\alpha=\left\{+,-\right\}}
      N_{\bk,\alpha}  \epsilon_\bk \right]
      +g n_+ n_-~,
\end{equation}
where $N_{\bk,\alpha}$ is a Fermi occupation factor.
The contribution to the energy density that is
second order in interactions is given by \cite{bookpathria}
\begin{eqnarray}
\label{eq:e2nd}
  e^{(2)} = - \frac{2 g^2}{V^3}  {\sum}'
  \frac{N_{\bk_1,+} N_{\bk_2,-}
  \left(N_{\bk_3,+}+N_{\bk_4,-}\right)}
  {\epsilon_{\bk_1}+\epsilon_{\bk_2}-\epsilon_{\bk_3}-\epsilon_{\bk_4}}~,
\end{eqnarray}
where the prime indicates that the sum is over wave vectors such
that $\bk_1+\bk_2=\bk_3+\bk_4$. The above second order correction
takes into account the so-called unitarity limit, {\em i.e.}, the
energy dependence of the vacuum scattering amplitude to all orders
in $ka$, to second order.

\noindent {\it Results} ---  The magnetization results, summarized
in Fig.~\ref{fig:combined}, were obtained by numerically
minimizing the total free energy $f(n_-,n_+)$ {\em vs.} $\xi
\equiv (n_+-n_-)/(n_+ + n_-)$, for a series of temperatures and
total densities $n_++n_-=k_{\rm F}^3/3 \pi^2$.  At zero
temperature, we find that the system becomes partially polarized
if $k_{\rm F} a \geq 1.054$, and reaches the fully-polarized state
at $k_{\rm F} a = 1.112$. For higher temperatures interactions
have to be stronger to polarize the system.  For temperatures
$T<T_{\rm tc}$, where $T_{\rm tc} \simeq 0.2 T_{\rm F}$ with
$T_{\rm F}$ the Fermi temperature, the transition is
discontinuous, and the magnetization exhibits a jump. The jump
becomes smaller with increasing temperature, vanishing at $T_{\rm
tc}$. The inset shows the transition temperature as a function of
$k_{\rm F}a$. A line of first-order transitions, denoted by the
solid line, joins a line of continuous transitions, denoted by the
dotted line at $T=T_{\rm tc}$ and $k_{\rm F} a = 1.119$.

The first order behavior at low temperatures is expected on the
basis of the arguments of Belitz {\it et al.} \cite{belitz1999}.
In our case the gapless modes that drive the transition first
order are particle-hole excitations. The coupling of these
excitations to the magnetization is neglected in Hartree-Fock
theory, which therefore always predicts a continuous transition.
Eq.~(\ref{eq:e2nd}) takes the coupling between the magnetization
and the particle-hole excitations into account to lowest order.

\begin{figure}
\includegraphics{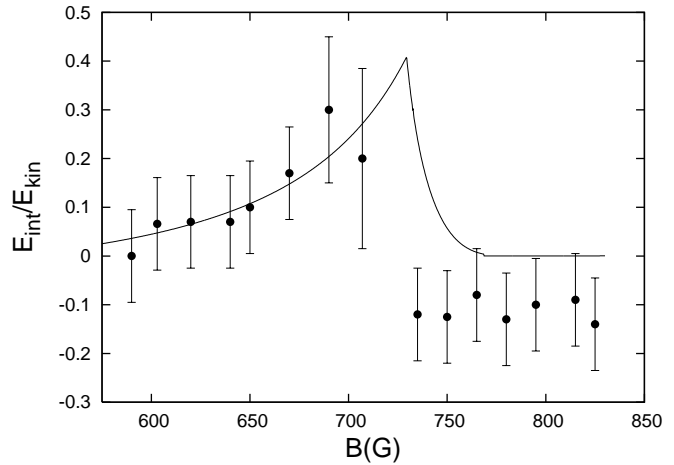}
\caption{\label{fig:ratio} Interaction energy divided by
kinetic energy as a function of magnetic field for the
experimental parameters of Bourdel {\it et al.}
\cite{bourdel2003}. We take a temperature $T=3.5$ $\mu$K $=0.6
T_{\rm F}$. For details on the magnetic-field dependence of the
scattering length see, for example, Ref.~\cite{bourdel2003}. }
\end{figure}

\noindent {\it Experimental implications} --- The ferromagnetic
state can be identified by measuring the interaction energy,
either by studying the expansion properties of the gas
\cite{bourdel2003}, or by using RF spectroscopy
\cite{chin2004,gupta2003,regal2003,duine2004}. The fully polarized
state is distinguished by the absence of any interaction energy.
In the experiments by Bourdel {\it et al.} \cite{bourdel2003} on
$^6$Li gases, the interaction energy appears to vanish when the
regime of strong repulsive interactions is approached. In
Fig.~\ref{fig:ratio} we plot the interaction energy divided by the
kinetic energy for their experimental parameters, as a function of
the magnetic field. Given the fact that we have not taken into
account the inhomogeneity of the system, the agreement is
remarkable, strongly suggesting that a ferromagnetic transition
occurs in this system. If we interpret the experimental data
accordingly, the transition is found to occur at $k_{\rm F} a
\simeq 1$ at $T=0.6 T_{\rm F}$, which is slightly smaller than our
calculated value ($k_{\rm F} a = 1.56$) at this temperature. In
the experiments of Gupta {\it et al.} \cite{gupta2003} the value
of $k_{\rm F} a$ at which the mean-field shift appears to vanish
is even smaller compared to the value we predict for the onset of
ferromagnetism. Since the atom system in these experiments is
prepared in a ferromagnetic state, these discrepancies between
theory and experiment could be due to the rapid increase in spin
coherence time which is expected as stable ferromagnetism is
approached, as we now explain.

Pseudospin decoherence in these systems is due to spatial
inhomogeneities in the Zeeman energy $\Delta E$.  Suppose the potential the
$|\!\uparrow\rangle$ atoms feel is $E^\uparrow (\bx)$, and the
potential the $|\!\downarrow\rangle$ atoms feel is $E^{\downarrow}
(\bx)$. The decay rate of the fully coherent state is suppressed
because the quasiparticle energies of the unoccupied pseudospins
are shifted by the interactions. Fermi's golden rule implies a coherence
decay rate
\begin{equation}
 \Gamma = \frac{2 \pi}{\hbar} \sum_{\bk',\bk}
 \left|\Delta E_{\bk',\bk}\right|^2 \delta(\epsilon_\bk-\epsilon_{\bk'}-g n)
 \left( N_{\bk,+} - N_{\bk',-}\right)~,
\end{equation}
where
\begin{equation}
 \Delta E_{\bk',\bk} = \frac{1}{V} \int d \bx \left( \frac{E^{\uparrow} (\bx)-E^\downarrow (\bx)}{2}
 \right) e^{i (\bk'-\bk)\cdot \bx}~.
\end{equation}
The consequences of interactions can be illustrated by taking
$ |\Delta E_{\bk',\bk}|^2 = \delta E^2 e^{-\Lambda^2
  (\bk-\bk')^2}$
where $\Lambda$ is the length scale of magnetic-field
inhomogeneities. In Fig.~\ref{fig:suppression} the $T=0$ spin
coherence time is shown for a series values of $\Lambda$.  These
results were obtained by taking a trapping frequency $\omega/2\pi
= 20$ Hz , estimating $\delta E$ as the difference in Zeeman
splitting change between the edge and center of the cloud, and
assuming $7 \times 10^{6}$ atoms at a density $n \sim 4 \times
10^{13}$ cm$^{-3}$, following Gupta {\em et al.} \cite{gupta2003}.
Clearly, the spin coherence time is strongly enhanced for
increasing interactions. The difference in experimental results
between Bourdel {\it et al.} \cite{bourdel2003} and Gupta {\it et
al.} \cite{gupta2003} might be related to differences in magnetic
field.  We note that the magnetic-field inhomogeneities are
necessary for the equilibration of the hyperfine spin degrees of
freedom. Since molecule formation cannot occur in the fully
polarized state, ferromagnetism competes kinetically with
Bose-Einstein condensation of molecules \cite{nygaard}.

Coherence decay and atomic ferromagnetism can
also be studied by measuring the size of the cloud.
In a local density approximation, valid since the
oscillator length exceeds the Fermi wavelength,
the size of the cloud is proportional to the square root of
the Fermi energy. It follows that the radius of the
fully-polarized state is a factor $2^{1/3}$ larger than that of the
unpolarized state.  The first order character of the phase transition
could be detected by performing
experiments with a mixture of fermions and bosons.
(The interactions between bosons and the fermions should be
weak enough to make boson mediated attractive interactions between
fermions negligible.) Suppose for example that the mixed system
is in equilibrium, and that the energy and number of atoms are
conserved as the bias field is varied.  Adiabatically increasing $k_{\rm F} a$ from the
paramagnetic to the ferromagnetic regime will lead to a temperature
increase that is tied to the entropy reduction in the ordered state.
For $T < T_{tc}$ the temperature variation should be hysteretic.
These temperature changes, although typically relatively small ($ \sim 10^{-3} T_{\rm
F}$), are larger for a smaller boson to fermion mass ratio and boson concentration
and might be observable.

\noindent {\it Discussion and conclusions.} --- Although
ferromagnetism is a strong coupling instability and our theory is
perturbative, we nevertheless believe that the phase-diagram in
the inset of Fig.~\ref{fig:combined} is reliable. The interaction
energy of the fully-polarized state, which is an eigenstate of the
full hamiltonian, is exactly zero. Moreover, a calculation to
third order in the gas parameter shows that the energy of the
paramagnetic state energy is increased in comparison to the
second-order result \cite{dedominicis1957}. Hence we expect that
the second-order perturbation theory underestimates the transition
gas parameter. However, since consistency requires that the
critical point lies in the strong-coupling regime where $k_{\rm F}
a \sim 1$, there appears to be little room for movement. An
experimental determination of the phase diagram appears to be
within reach and would be interesting. The magnetic properties of
ultracold Fermi gases could provide a very interesting system to
explore fundamental aspects of ferromagnetism, including the
dynamics of domains walls which could be directly manipulated by a
one-way barriers \cite{raizen2005}, the nonequilibrium formation
of the ferromagnetic state, spin waves, and spin transfer effects.

It's a pleasure to thank Deep Gupta, Randy Hulet, Michael Jack,
Wolfgang Ketterle, Nicolai Nygaard, Jamie Williams, and Martin
Zwierlein for discussions. This work was supported by the National
Science Foundation under grant DMR-0115947 and by the Welch
Foundation.

\end{document}